# A new M-matrix of Type III, its properties and applications

R.N.Mohan[1], Moon Ho Lee[2], and Ram Paudal[3].
Sir CRR Institute of Mathematics, Eluru-534007, AP, India[1];
Institute of Information & Communication, Chonbuk National University, Korea[2,3].
Email: mohan420914@yahoo.com,[1], moonho@chonbuk.ac.kr [2]

**Abstract:** Some binary matrices like (1,-1) and (1,0) were studied by many authors like Cohn [3], Wang [31], Ehlich [5] and Ehlich and Zeller [6], and Mohan, Kageyama, Lee, and Gao in [18]. In this recent paper by Mohan et al considered the M-matrices of Type I and II by studying some of their properties and applications. In the present paper we discuss the M-matrices of Type III, and study their properties and applications. We give some constructions of SPBIB designs and some corresponding M-graphs, which are being constructed by it. This is the continuation of our earlier research work in this direction, and these papers establish the importance of non-orthogonal matrices as well.

**1. Introduction.** There are numerous types of matrices, and among them (1,-1) matrices and (0,1) called as binary matrices, which were constructed and studied by Cohn [3], Wang [31], Ehrlich [5] and Ehrlich and Zeller[6]. They play pivotal role in signal and image processing [14], and in the construction of codes, designs, graphs, refer to Colbourn, Dinitz, and Stinson [4], and sequences and array sequences by Fan and Darnell [7]. Specifically Ehlich [5], Ehlich and Zeller [6], Seifer [25] and Wang [31] studied these binary matrices and their properties and Kahn, Kolmos and Szemeredi [10] studied the probability of these matrices to be singular. Again the well known Hadamard matrix, which is also a (1,-1)-matrix has orthogonal property for details refer to Geramita and Seberry [8]. And Seberry and Yamada [26] has shown many such applications. . These matrices have numerous applications in the construction of codes, designs, and graphs [8,26]. For applications in signals and sequences refer to Fan and Darnell [7]. These matrices are extensively used in the image reconstruction in image processing (Teague [29]). Mohan [16] defined an $M_n$-matrix $=\left[ a_{ij} \right]$, by using $a_{ij} = (d_i \otimes d_h d_j) \bmod n$ and by suitably defining $d_i$, $d_h$, $d_j$ and $\otimes$, which is given below.

**Definition 1.1.** *The $M_n$-matrix can be defined as a matrix for n being a prime and as a matrix obtained from $\left[ a_{ij} \right]$ where $a_{ij} = 1 + (i-1)(j-1) \bmod n$, $i, j = 1, 2, ..., n$. This is an $n \times n$ symmetric matrix.*

The pattern of this matrix is being used in the construction of Low-density parity check (LDPC) code by Vasic and Olgica in [30].

As an extension of this concept, M-matrices of Type I and II have been defined by Mohan [16] and Mohan, Kageyama, Lee, Yang [18] as follows:

**Definition 1.2.** *When n is a prime, consider the matrix of order n obtained by the equation $M_n = [a_{ij}]$, where $a_{ij} = 1+(i-1)(j-1) \bmod n$, and $i, j = 1, 2, ...n$. In the resulting matrix retain 1 as it is and substitute -1's for odd numbers and +1's for even numbers. (We can substitute 1 for odd numbers and -1 for even numbers; in that case change of sign occurs in its determinant). Let the resulting matrix M be called as **M-matrix of Type I**. In it each column and each row contain (n+1)/2 number of +1's and (n-1)/2 number of -1's. This is an n×n symmetric matrix.*

Again the $M_n$-matrix can be obtained by $a_{ij}$= (i.j) mod n, when n is a prime and by extending this concept we have the M-matrix of type II as follows:

**Definition 1.3.** *The M-matrix of Type II is obtained by the equation $a_{ij} = (i.j) \bmod (n+1)$, where n+1 is a prime. In this matrix since each row and each column have n elements and in all the columns and rows from 1 to n elements, each element comes once in each row and each column. In the resulting matrix substitute 1 for even numbers and -1 for odd numbers and also for 1, ( or 1 for odd numbers keeping the 1 in the matrix as 1 itself and -1 for even numbers). Then this resulting matrix M is called the **M-matrix of Type II**. When n is odd there are the number of +1's is (n+1)/2 and the number of -1's is (n-1)/2. When n is even each row (column) consists of equal number of +1's and -1's are equal to n/2. This is also an n×n symmetric matrix.*

## 2. Main Results

As the day stands there are only three possible ways of taking $a_{ij} = (d_i \otimes d_h d_j) \bmod n$, when n is prime by suitably defining $d_i, d_h, d_j$, and $\otimes$ as follows:

1. $a_{ij} = 1+(i-1)(j-1) \bmod n$, when n is prime.
2. $a_{ij} = (i.j) \bmod n$, when n is prime.

(We considered n+1 as a prime for our need here).

3. $a_{ij} = (i+j) \bmod n$, when n is a positive integer.

The first two types have already been studied by Mohan, et al in [18], and in this paper we propose to explore the third possibility for n as an integer (both for odd and even cases). Note that these three are combinatorially equivalent but structurally different and hence need to be studied independently. These non-orthogonal matrices are considered not of much importance so far but now we propose to show the need of these types of matrices. The non-orthogonality is being used in image reconstruction refer to Mathematica [14] in image processing.

Again the $M_n$-matrix can be obtained by $a_{ij}$= (i+j) mod n, when n is a prime. By extending this concept we have the M-matrix of Type III as follows:

**Definition 2.1.** The **M-matrix of Type III** is obtained by the equation $a_{ij} = (i+j) \bmod n$, where n is odd or even number. In this matrix since each row or column has n elements. when n is even, 1 to n elements do come in all the columns and rows, and each element comes once in each row and each column. In the resulting matrix substitute 1 for even numbers and -1 for odd numbers and also for 1, ( or 1 for odd numbers keeping the 1 in the matrix as 1 itself and -1 for even numbers). Then this resulting matrix M is called M-matrix of Type III. When n is odd, in each row and each column the number of +1's is (n+1)/2 and the number of -1's is (n-1)/2. When n is even in each row (column) consists of equal number of +1's and -1's numbering to n/2. This is also an n×n symmetric matrix.

When n is odd then the elements of the principal diagonal of the Type III has a special property, i.e. if $D_n$ is the principal diagonal of the M-matrix of order n, then we get its elements by (2x) mod n. Thus for example $D_3$ = (1 2 3), $D_5$ = (1 2 3 4 5) and $D_9$ = (1 2 3 4 5 6 7 8 9). All elements occur due to modular property. In the case when n is even (2x) mod n, the elements in the principal diagonal will be half of them repeat again, all are even numbers only. For example $D_4$ = (2 4 2 4), $D_6$ = (2 4 6 2 4 6) and so on.

With the help of these matrices many SPBIB designs have been constructed. For details regarding PBIB and SPBIB designs refer to Raghavarao [20,21], Bose [1,2], Kageyama and Mohan [11], Liu [13], Kusumoto [12], Mohan [15], Ogasawara [19], Ramanujacharyulu [22], Roy [23,24], Shrikhande [27], and Sprott [28].

**Proposition 2.1.** In a given M-matrix of type III when n is odd, in each of its rows and columns, the number of +1's are (n+1)/2 and the number of -1's are (n-1)/2.

**Proof.** Since n is odd in each row (column) each element of 1, 2, 3, …, n occurs exactly once. And among these (n+1) elements are odd numbers (n-1)/2 are even numbers. Consequently as we replace even numbers by -1's and odd numbers by +1's, retaining the +1 as it is we get that (n+1)/2 elements are +1's and (n-1)/2 elements are -1's.□

As these M-matrices are non-orthogonal, we define the orthogonal numbers for them as follows:

**Definition 2.1.** The **orthogonal number** of a given M-matrix with entries ± 1 is defined as sum of the products of the corresponding numbers in two given rows of the matrix (called inner product of the rows). Consider any two rows $R_l = (r_1, r_2, ..., r_n)$ and $R_m = (s_1 s_2, ..., s_n)$ and then the orthogonal number denoted by 'g' can be defined as $g = (R_l R_m) = \sum_{i=1}^{n} r_i s_i$.

**Proposition 2.2.** In an M-matrix of Type III, when n is odd, the orthogonal number between any two rows $R_i$ and $R_j$ where $i \neq j$ is given by 4k-2-n, where k is the number of unities in the selected set, where $1 \leq k \leq \frac{n+1}{2}$.

**Proof.** Let $R_i$, and $R_j$ be the two given rows in the M-matrix. We have to calculate their inner product $\langle R_i, R_j \rangle$, where $i \neq j$ and $i \neq 1$. By elementary transformations we can make the row $R_i$ with the first $\frac{n+1}{2}$ elements as +1's and the next $\frac{n-1}{2}$ elements as -1's. The orthogonal numbers of the matrix remains invariant by such elementary transformations. Now consider the other row $R_j$, which has n elements. Divide them in to two sets, the first with (n+1)/2 elements and the second with $\frac{n-1}{2}$ elements. They can be depicted as follows:

$$R_i = \underbrace{(1\,1\,1\,1\ldots1\,1\,1\,1\ldots1)}_{\frac{n+1}{2}\text{ elements}} \quad \underbrace{(-1\,-1\,-1\,-1\ldots-1\,-1\,-1\,-1)}_{\frac{n-1}{2}\text{ elements}}$$

$$R_j = \underbrace{(1\,-1\,-1\ldots-1\,-1\ldots1\,1)}_{\text{let +1's be } k \text{ in number hence -1's are } \frac{n+1}{2}-k} \quad \underbrace{(1\,1\ldots1\,-1\,1\,-1\,-1\,-1\,-1\,1\,1\ldots1\,1)}_{+1\text{'s are } \frac{n+1}{2}-k \text{ and } -1\text{'s are } (k-1) \text{ in number}}$$

Now we evaluate the formula for the orthogonal number. Let k be the number of +1's. In the first two sets of the given rows (i) the k number of +1's of the row $R_j$ correspond with k number of +1's in the row $R_i$. And (n+1)/2 –k number of -1's of the row $R_j$ coincide with same number of +1's of the row $R_i$ in the second sets of the two given rows (iii) $\frac{n+1}{2}$ -k number of +1's of the row $R_j$ correspond with the same number of -1's of the row $R_i$ and (iv) the k-1 number of −1's of the row $R_j$ correspond with same of -1's in the row $R_i$. Hence we get the orthogonal number g as

$$g = \langle R_i, R_j \rangle = k \times 1 \times 1 + ((\frac{n+1}{2} - k) \times 1 \times (-1)) + ((\frac{n+1}{2} - k) \times (+1) \times (-1)) + (k-1)(-1)(-1) = 4k - 2 - n.$$

If we see for different values of k, it is not possible that k = 0, since the first set of $R_j$ can not have (n+1) number of +1's. For k =1, g = 2-n, for k=2, g = 6-n, and so on. And when k = (n+1)/2 then g = n. There are (n+1)/2 orthogonal numbers in the M-matrix of type III. □

**Result 2.1.** We have $\langle R_i, R_i \rangle = n, where\ i = 0, 1, 2, \ldots, n$.

**Proof.** For $\langle R_i, R_i \rangle = n$, the inner product any row with itself, gives out n only. Besides when k = $\frac{n+1}{2}$, then also we get $\langle R_i, R_j \rangle = n$. That is called the trivial orthogonal number.□

**Note 2.1.** In the M-matrix there are some orthogonal numbers which occur in pairs, which are called as orthogonal pairs. In a pair the first orthogonal number occurs if the number of

+1's is $\theta_1$ and the number of -1's is $\theta_2$. And the second orthogonal number occurs if the number of -1's is $\theta_1$ and the number of +1's is $\theta_2$, where $\theta_1 + \theta_2 = \dfrac{n+1}{2}$. The sum of the orthogonal numbers $g_i$ and $g_j$ of an orthogonal pair $= g_i + g_j = -2$. Proof is trivial.□

**Proposition 2.3.** In this M-matrix the sum of the orthogonal numbers is $\dfrac{n+1}{2}$.

**Proof.** When $g_i$'s are orthogonal numbers of the given M-matrix of Type III, then $\sum\limits_{i=1}^{\frac{n+1}{2}} g_i = \sum\limits_{k=1}^{\frac{n+1}{2}} 4k - 2 - n = (n+1)/2$, which is an arithmetic progression with AM as 4. Hence the proof.□

**Result 2.2.** For the given M-matrix $|M| = (-1)^{\frac{n-1}{2}} 2^{n-1}$.

Proof: Consider sum of any two rows in the matrix, which is a vector with all zeros except one element that comes out to be 2. Thus if we reduce the matrix with $R_i + R_{i+1}$, where i = 1, 2, …, n-1, we get the first n-1 rows with all (n-1)zeros and one 2 in different columns. Then considering the determinant of this matrix take each row, which is the product of one 2 and the remaining matrix. Note the last row remains as it is with due reduction. Then lastly the determinant will be 2 only. Thus it is a product of $(-1)^{\frac{n-1}{2}} 2^{n-1}$, as in the n-1 number of 2, half with + sign and the remaining half with - sign.□

**Example 2.1.** We take when n is odd. Let n = 3. Consider (i+j) mod n, for i, j = 1, 2, …, n

Then we get the following $M_n$-matrix.

$$\begin{bmatrix} 2 & 3 & 1 \\ 3 & 1 & 2 \\ 1 & 2 & 3 \end{bmatrix}.$$

By substituting -1's for even numbers and +1's for odd numbers and keeping 1 in the matrix as it is we get the M-matrix of type III as

$$\begin{bmatrix} -1 & 1 & 1 \\ 1 & 1 & -1 \\ 1 & -1 & 1 \end{bmatrix}.$$

The orthogonal number is -1 only and the trivial orthogonal number is 3.

Treating -1's as 0's we get the incidence matrix of a design as follows:

$$\begin{bmatrix} 0 & 1 & 1 \\ 1 & 1 & 0 \\ 1 & 0 & 1 \end{bmatrix}.$$

This yields an SBIB design (3, 2, 1). In the case of Type I, we get another type of design and the structure is different.

**Example 2.2.** Take n = 5. Consider i, j = 1, 2, 3, 4, 5. The equation (i+j) mod n gives the following $M_n$-matrix.

$$\begin{bmatrix} 2 & 3 & 4 & 5 & 1 \\ 3 & 4 & 5 & 1 & 2 \\ 4 & 5 & 1 & 2 & 3 \\ 5 & 1 & 2 & 3 & 4 \\ 1 & 2 & 3 & 4 & 5 \end{bmatrix}.$$

By substituting -1 for even numbers and +1 for odd numbers and keeping 1 in the matrix as it is we get the M-matrix of Type III.

$$\begin{bmatrix} -1 & 1 & -1 & 1 & 1 \\ 1 & -1 & 1 & 1 & -1 \\ -1 & 1 & 1 & -1 & 1 \\ 1 & 1 & -1 & 1 & -1 \\ 1 & -1 & 1 & -1 & 1 \end{bmatrix}.$$

The orthogonal numbers are given by $\langle R_i, R_j \rangle$ as follows:
$\langle R_1, R_2 \rangle = \langle R_2, R_3 \rangle = \langle R_3, R_4 \rangle = \langle R_4, R_5 \rangle = -3$
$\langle R_1, R_3 \rangle = \langle R_1, R_4 \rangle = \langle R_1, R_5 \rangle = \langle R_2, R_5 \rangle = \langle R_3, R_5 \rangle = 1$
$\langle R_i R_i \rangle = 5$, which is trivial orthogonal number.

By taking -1's as 0's we get incidence matrix of a design.

$$\begin{bmatrix} 0 & 1 & 0 & 1 & 1 \\ 1 & 0 & 1 & 1 & 0 \\ 0 & 1 & 1 & 0 & 1 \\ 1 & 1 & 0 & 1 & 0 \\ 1 & 0 & 1 & 0 & 1 \end{bmatrix}.$$

This yields an SPBIB design with parameters v=b= 5, r=k= 3, $\lambda_1$= 1, $\lambda_2$ = 2, $n_1$=2, $n_2$ = 2. In case of Type I we get a different design as it is a different structure.

$$P_1 = \begin{bmatrix} p_{11}^1 = 0 & p_{12}^1 = 1 \\ p_{21}^1 = 1 & p_{22}^1 = 1 \end{bmatrix}, P_2 = \begin{bmatrix} p_{11}^2 = 1 & p_{12}^2 = 1 \\ p_{21}^2 = 1 & p_{22}^2 = 0 \end{bmatrix}.$$

**Note 2.2.** In these two examples even though 3 and 5 are primes here the structures are not like that of Type I, as these are obtained from (i+j) mod n.

**Note 2.3.** In the case when n is even then though the structure is different to that of M-matrix of Type II, the formula for the orthogonal number remains the same g = 4k-n, as given in [18], and hence can be dealt with in the similar manner, consequently it has been omitted from the present discussions. But the subtleties in this case will be considered in our further research in this direction.

**Example 2.3.** Take n = 9. Then from the equation (i+j) mod n we get the $M_n$-matrix as

|   | 1 | 2 | 3 | 4 | 5 | 6 | 7 | 8 | 9 |
|---|---|---|---|---|---|---|---|---|---|
| 1 | 2 | 3 | 4 | 5 | 6 | 7 | 8 | 9 | 1 |
| 2 | 3 | 4 | 5 | 6 | 7 | 8 | 9 | 1 | 2 |
| 3 | 4 | 5 | 6 | 7 | 8 | 9 | 1 | 2 | 3 |
| 4 | 5 | 6 | 7 | 8 | 9 | 1 | 2 | 3 | 4 |
| 5 | 6 | 7 | 8 | 9 | 1 | 2 | 3 | 4 | 5 |
| 6 | 7 | 8 | 9 | 1 | 2 | 3 | 4 | 5 | 6 |
| 7 | 8 | 9 | 1 | 2 | 3 | 4 | 5 | 6 | 7 |
| 8 | 9 | 1 | 2 | 3 | 4 | 5 | 6 | 7 | 8 |
| 9 | 1 | 2 | 3 | 4 | 5 | 6 | 7 | 8 | 9 |

**Note 2.3.** In the three examples above it may be observed that these are circulant matrices, which are defined as n x n matrix whose rows are composed of cyclically shifted versions of a length *n* and a list *l*. And the list may consist of any elements like ($a_1, a_2, \ldots, a_n$) on which no property was defined. But the matrix defined here is governed by (i+j) mod n, where n is odd or even and i,j = 1,2,…, n. But these types of matrices are very useful in digital image processing [14].

Now substituting -1 for even numbers and +1 for odd number and keeping 1 as it is we get the M-matrix of the third type as

|   | 1 | 2 | 3 | 4 | 5 | 6 | 7 | 8 | 9 |
|---|---|---|---|---|---|---|---|---|---|
| 1 | -1 | 1 | -1 | 1 | -1 | 1 | -1 | 1 | 1 |
| 2 | 1 | -1 | 1 | -1 | 1 | -1 | 1 | 1 | -1 |
| 3 | -1 | 1 | -1 | 1 | -1 | 1 | 1 | -1 | 1 |
| 4 | 1 | -1 | 1 | -1 | 1 | 1 | -1 | 1 | -1 |
| 5 | -1 | 1 | -1 | 1 | 1 | -1 | 1 | -1 | 1 |
| 6 | 1 | -1 | 1 | 1 | -1 | 1 | -1 | 1 | -1 |
| 7 | -1 | 1 | 1 | -1 | 1 | -1 | 1 | -1 | 1 |
| 8 | 1 | 1 | -1 | 1 | -1 | 1 | -1 | 1 | -1 |
| 9 | 1 | -1 | 1 | -1 | 1 | -1 | 1 | -1 | 1 |

The orthogonal number are given by

$\langle R_i R_i \rangle = 9$, which is the trivial orthogonal number.

$\langle R_1, R_2 \rangle = \langle R_1, R_9 \rangle = \langle R_2, R_3 \rangle = \langle R_3, R_4 \rangle = \langle R_4, R_5 \rangle = \langle R_5, R_6 \rangle = \langle R_6, R_7 \rangle = \langle R_7, R_8 \rangle = \langle R_8, R_9 \rangle = -7$

$\langle R_1, R_4 \rangle = \langle R_1, R_7 \rangle = \langle R_2, R_5 \rangle = \langle R_2, R_8 \rangle = \langle R_4, R_7 \rangle = \langle R_5, R_7 \rangle = \langle R_5, R_8 \rangle = \langle R_6, R_9 \rangle = -3$

$\langle R_1, R_5 \rangle = \langle R_1, R_6 \rangle = \langle R_2, R_6 \rangle = \langle R_2, R_7 \rangle = \langle R_3, R_6 \rangle = \langle R_3, R_7 \rangle = \langle R_3, R_8 \rangle = \langle R_3, R_9 \rangle = \langle R_4, R_8 \rangle = \langle R_4, R_9 \rangle = \langle R_5, R_9 \rangle = 1$

$\langle R_1, R_3 \rangle = \langle R_1, R_8 \rangle = \langle R_2, R_4 \rangle = \langle R_2, R_9 \rangle = \langle R_3, R_5 \rangle = \langle R_4, R_6 \rangle = \langle R_6, R_8 \rangle = \langle R_7, R_9 \rangle = 5$.

In this M-matrix substitute 0's for -1's we get

|   | 1 | 2 | 3 | 4 | 5 | 6 | 7 | 8 | 9 |
|---|---|---|---|---|---|---|---|---|---|
| 1 | 0 | 1 | 0 | 1 | 0 | 1 | 0 | 1 | 1 |
| 2 | 1 | 0 | 1 | 0 | 1 | 0 | 1 | 1 | 0 |
| 3 | 0 | 1 | 0 | 1 | 0 | 1 | 1 | 0 | 1 |
| 4 | 1 | 0 | 1 | 0 | 1 | 1 | 0 | 1 | 0 |
| 5 | 0 | 1 | 0 | 1 | 1 | 0 | 1 | 0 | 1 |
| 6 | 1 | 0 | 1 | 1 | 0 | 1 | 0 | 1 | 0 |
| 7 | 0 | 1 | 1 | 0 | 1 | 0 | 1 | 0 | 1 |
| 8 | 1 | 1 | 0 | 1 | 0 | 1 | 0 | 1 | 0 |
| 9 | 1 | 0 | 1 | 0 | 1 | 0 | 1 | 0 | 1 |

By taking this matrix as an incidence matrix of design, yields an SPBIB design with parameters v = b = 9, r = k = 5, $\lambda_1$=1, $\lambda_2$ = 2, $\lambda_3$ = 3, $\lambda_4$ = 4, $n_1$=2, $n_2$ = 2, $n_3$ = 2, n = 2, which is a 4-associate class SPBIB design.

$$P_1 = \begin{bmatrix} p^1_{11}=0 & p^1_{12}=0 & p^1_{13}=0 & p^1_{14}=1 \\ p^1_{21}=0 & p^1_{22}=0 & p^1_{23}=1 & p^1_{14}=1 \\ p^1_{31}=0 & p^1_{32}=1 & p^1_{33}=1 & p^1_{14}=0 \\ p^1_{41}=1 & p^1_{42}=1 & p^1_{43}=0 & p^1_{44}=0 \end{bmatrix}, P_2 = \begin{bmatrix} p^2_{11}=0 & p^2_{12}=0 & p^2_{13}=1 & p^2_{14}=1 \\ p^2_{21}=0 & p^2_{22}=1 & p^2_{23}=0 & p^2_{14}=0 \\ p^2_{31}=1 & p^2_{32}=0 & p^2_{33}=0 & p^2_{14}=1 \\ p^2_{41}=1 & p^2_{42}=0 & p^2_{43}=1 & p^2_{44}=0 \end{bmatrix},$$

$$P_3 = \begin{bmatrix} p^3_{11}=0 & p^3_{12}=1 & p^3_{13}=1 & p^3_{14}=0 \\ p^3_{21}=1 & p^3_{22}=0 & p^3_{23}=0 & p^3_{14}=1 \\ p^3_{31}=1 & p^3_{32}=0 & p^3_{33}=0 & p^3_{14}=0 \\ p^3_{41}=0 & p^3_{42}=1 & p^3_{43}=0 & p^3_{44}=1 \end{bmatrix}, P_4 = \begin{bmatrix} p^4_{11}=1 & p^4_{12}=1 & p^4_{13}=0 & p^4_{14}=0 \\ p^4_{21}=1 & p^4_{22}=0 & p^4_{23}=1 & p^4_{14}=0 \\ p^4_{31}=0 & p^4_{32}=1 & p^4_{33}=0 & p^4_{14}=1 \\ p^4_{41}=0 & p^4_{42}=0 & p^4_{43}=1 & p^4_{44}=0 \end{bmatrix}.$$

The construction graphs by PBIB designs had been done by Bose [2], Goethels and Siedel [9]. Now the graph constructed by the above binary matrix is called M-graph.

By treating that the above binary matrix as an adjacency matrix of a graph we get the M-graph as follows:

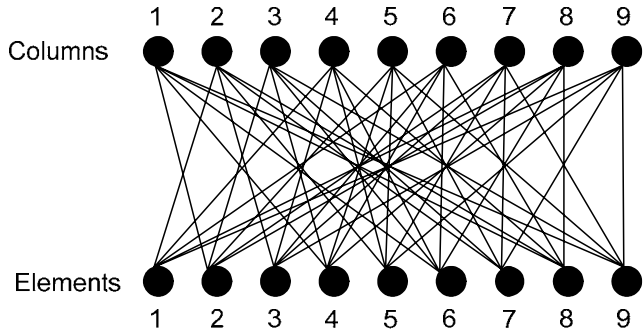

This type of M-graphs are found to be highly fault tolerant, richly connected architectures for being a Network system usable in multi-processor and communication systems.

For more application of Combinatorial designs refer to Colbourn, Dinitz and Stinson [4]. For technical aspects of this M-network refer to Mohan and Kulkarni [17] and a sequel to this paper to appear soon.

**Acknowledgements:** One of the authors Mohan is thankful to Prof. M. G. K. Menon, who is a fountainhead of inspiration to him and to the Third World Academy of Sciences, Trieste, Italy and Prof. Bill Chen, Center for Combinatorics (CFC), Nankai University, Tianjin, PR China, for giving him an opportunity to work in the center in China for three months. His thanks are also due to Sir CRR College authorities namely K.Srimanarayana, Principal and Gutta Subbarao, Secretary for their kind support in his research quest. He is also thankful to Prof. Moon Ho Lee for extending invitation to visit Chonbuk National University, South Korea.This has ben modified during Mohan's second visit i.e., March 21- June 19, 2006, to CFC, PR China.

This work was partially supported by the MIC (Ministry of Information and Communication), under the ITFSIP (IT Foreign Specialist Inviting Program ) supervised by IITA, under ITRC supervised by IITA, and International Cooperation Research Program of the Ministry of Science & Technology, Chonbuk National University, Korea and partially by the Third World Academy of Sciences, Italy, Hence all the concerned authorities are gratefully acknowledged.

**References**

[1] Bose, R.C., and Nair, K.R. (1939). Partially balanced incomplete block designs. *Sānkhya*, Vol. **4,** pp.337-372.

[2] Bose, R.C. (1963). Strongly regular graphs, partial geometries and partially balanced designs. Pacific J. Math., Vol. **13,** pp. 389-419.

[3] Cohn, J.H.E. (1963). Determinants with elements $\pm 1$. J.London Math.Soc., Vol.**14**, 581-588.

[4] Colbourn, C.J., Dinitz, J.H., and Stinson, D.R. (1993,1999) Applications of Combinatorial Designs to Communications, Cryptography, and Networking (1999)**.** Surveys in Combinatorics, 1993, Walker (Ed.), London Mathematical Society Lecture Note Series 187, Cambridge University Press.

[5] Ehlich, H. (1964). Determinantenabschätzungen für binäre Matrizen. Math. Z., Vol. **83,** 123-132.

[6] Ehlich, H., and Zeller, K. (1962). Binäre Matrizen. Z. Angew. Math. Mech., Vol. **42,** T 20-21.

[7] Fan, P., and Darnell, M. (1996). *Sequence design for communications applications.* Research Studies Press Ltd. John Wiley & Sons Inc.

[8] Geramita, Anthony,V., and Sebbery, Jennifer. (1979). *Orthogonal designs, Quadratic forms, and Hadamard matrices.* Marcel and Dekker Inc., New York.

[9] Goethals, J.M. and Siedel, J.J. (1970). Strongly regular graphs derived from combinatorial designs. Can. J. Math., Vol. **22**, pp. 597-614.

[10] Kahn,J., Komlos,J., and Szemeredi, E. (1995). On the probability that a Random $\pm$matrix is singular. J.Amer.Math.Soc., Vol. **8**, pp. 223-240.

[11] Kageyama Sanpei and Mohan, R. N. (1984) On the Construction of group divisible partially balanced incomplete block designs. Bull. of the Faculty of School Education, Hiroshima University, Part II, Vol. **7**, pp. 57-60.

[12] Kusumoto, K. (1967). Association schemes of new types and necessary conditions for the existence of regular and symmetrical PBIB designs with those association schemes. Ann.Inst.Statist.Math., Vol. **19,** pp.73-100.

[13] Liu, C.-W. (1963). A method of constructing certain symmetrical partially balanced designs. Scientia Sinica, Vol. **12**, pp.1935-1936.

[14] Mathematica Digital Image Processing.

[15] Mohan, R.N. (1999). A new series of μ-resolvable (d+1)-associate class GDPBIB designs. Indian J. Pure Appl.Math., Vol. **30**(1), pp. 89-98.

[16] Mohan, R.N. (2001). Some classes of $M_n$-matrices and $M_n$-graphs and their applications. JCISS., Vol. **26**, 1-4, pp.51-78.

[17] Mohan, R .N., and Kulkarni, P.T. (2006). A new family of fault-tolerant M-networks, (IEEE Trans. Computers, under revision).

[18] Mohan,R.N., Sanpei Kageyama, Moon Ho Lee, and Gao Yang. (2006) Certain new M-matrices and their properties and applications, Submitted to Linear Algebra and Applications, Released as e-print, arXiv: math.CS/0604035 Dt. April 9, 2006.

[19] Ogasawara, M. (1965). A necessary condition for the existence of regular and symmetrical PBIB designs of $T_m$ type. Inst.Statist. Mimeo. Series 418, Chapel Hill, N.C.

[20] Raghavarao, D. (1960). A generalization of group divisible designs. Ann.Math.Statist., Vol. **31**, pp.756-771.

[21] Raghavarao, D. (1971) *Constructions and Combinatorial problems in Design of Experiments.* Johan Wiley Inc., New York.

[22] Ramanujacharyulu, C. (1965). Non-Linear spaces in the construction of symmetric PBIB designs. *Sānkhya* ,Vol. **A27,** pp. 409-414.

[23] Roy, P.M. (1953-54). Hierarchical group divisible incomplete block designs with m-associate class. Science and Culture, Vol. **19**, pp. 210-211.

[24] Roy, P.M. (1962). On the properties and construction of HGD designs with m-associate classes. Calcutta Stat. Assn. Bull., Vol. **11,** pp.10-38.

[25] Siefter, N.(1984). Upper bounds for permanents of (1,-1)-matrices. Israel J. Math.

[26] Sebbery, Jennifer and Yamada, Mieko. (1992). *Hadamard Sequences, Block Designs. Contemporary design Theory:* A collection of Surveys, Edited by J.H.Dinitz, and D.R.Stinson, John Wiley & Sons, Inc., New York.

[27] Shrikhande, S.S. (1953). Cyclic solutions of symmetrical group divisible designs. Calcutta Statist.Assoc.Bull. Vol. **5,** pp.36-39.

[28] Sprott, D.A. (1959). A series of symmetric group divisible designs.Ann.Math.Statist.Vol. **30**, pp.249-251.

[29] Teague, M.R. (1979). Image analysis via the general theory of moments. J. Optical Soc. America, Vol.**70,** pp. 920-930.

[30] Vasic, Bane and Milenkovic, Olgica, (2004). Combinatorial constructions of low-density parity-check codes for iterative decoding. IEEE Trans. on Information Theory, Vol. **50**, No.6, pp.1156-1176.

[31] Wang,E.T.(1974). On permanents of (1,-1)-matrices. Israel J. Math., Vol. **18**, pp.353-361.